# Comparing Unit Trains versus Manifest Trains for the Risk of Rail Transport of Hazardous Materials - Part I: Risk Analysis Methodology


Di Kang[a], Jiaxi Zhao[b], C. Tyler Dick[b], Xiang Liu*[a], Zheyong Bian[c], Steven W. Kirkpatrick[d], Chen-Yu Lin[e]

[a] Department of Civil and Environmental Engineering, Rutgers, The State University of New Jersey, Piscataway, NJ.

[b] Rail Transportation and Engineering Center – RailTEC, Department of Civil and Environmental Engineering, University of Illinois at Urbana Champaign, Urbana, IL

[c] Department of Construction Management, University of Houston, Houston, TX.

[d] Applied Research Associates, Inc., Los Altos, CA.

[e] Department of Transportation and Logistics Management, National Yang Ming Chiao Tung University, Taipei, Taiwan.

**Corresponding Author:**

**Xiang Liu**, xiang.liu@rutgers.edu





**Abstract:** Transporting hazardous materials (hazmats) using tank cars has more significant economic benefits than other transportation modes. Although railway transportation is roughly four times more fuel-efficient than roadway transportation, a train derailment has greater potential to cause more disastrous consequences than a truck incident. Train types, such as unit train or manifest train (also called mixed train), can influence transport risks in several ways. For example, unit trains only experience risks on mainlines and when arriving at or departing from terminals, while manifest trains experience additional switching risks in yards. Based on prior studies and various data sources covering the years 1996-2018, this paper constructs event chains for line-haul risks on mainlines (for both unit trains and manifest trains), arrival/departure risks in terminals (for unit trains) and yards (for manifest trains), and yard switching risks for manifest trains using various probabilistic models, and finally determines expected casualties as the consequences of a potential train derailment and release incident. This is the first analysis to quantify the total risks a train may encounter throughout the shipment process, either on mainlines or in yards/terminals, distinguishing train types. It provides a methodology applicable to any train to calculate the expected risks (quantified as expected casualties in this paper) from an origin to a destination.

**Keywords:** risk analysis methodology; safety; freight railroads; hazardous material; mainlines; yard; terminal.


## 1. Introduction

The year 2021 has witnessed several train incidents on freight railways involving derailments and leaks of hazardous materials (hazmat): 47 cars of a train carrying combustible fertilizer and asphalt derailed in Sibley, Iowa; 27 cars derailed in Ames, Iowa; and 30 cars derailed in Newberry Township, Pennsylvania. The hazmat release poses a significant threat to surrounding people, property, and the environment. When transporting the same amount of hazmat using the same number of tank cars from the same origin to the same destination, service strategies play an important role in reducing the overall transportation risks. One possible strategy uses *unit trains*, usually consisting of 40-120 railcars, carrying the same commodity from the origin *terminal* to the destination terminal. Another possible service strategy uses *manifest trains*, in which railcars from multiple origins and destinations assemble and disassemble between trains at intermediate *classification yards*.

In the context of North American railroads, freight shipments carried by manifest trains require a process of assembling and dissembling, so that railcars bound for the same destination (or intermediate) classification yards re-sort into a new train. Railroad classification yards serve as hubs where loaded and empty railcars from various origins are grouped together into blocks of railcars headed for common destinations. These blocks of railcars are then further aggregated to form trains destined for different classification yards on other parts of the network or for local delivery to nearby shippers. The railcar sorting process requires numerous coupling and uncoupling events as groups of railcars are moved between multiple parallel tracks. In the highest-volume classification yards, sorting is accomplished by pushing railcars over small hills, or "humps." These events typically take place 1) in the main classification yard and its associated



tracks used for accumulating railcars into blocks by destination, 2) on the switching lead tracks used to actively sort the railcars and connect the receiving and departure tracks to the classification tracks, or 3) on other ancillary tracks used to process railcars as they pass through the classification yard. The sorting and switching process by destination in classification yards poses an additional risk of derailment and release.

In addition to the line-haul risk on mainlines, both unit and manifest trains encounter derailment and release risk during arrival and departure (A/D) events at loading and unloading terminals and classification yards. A/D events are operation processes similar to mainline operations but with a reduced speed on the non-mainline track. For unit trains at terminals, A/D events typically occur on loop or "balloon" tracks used to sequentially load or unload each railcar in the unit train as it advances at low speed or on the lead tracks connecting these facilities to the mainline. For manifest trains at classification yards, A/D events typically take place 1) in the receiving sub-yard where manifest trains arrive from the connection to the mainline, 2) in the departure sub-yard where manifest trains depart the classification yard to the mainline, or 3) on the lead and running tracks connecting the receiving and departure sub-yards to the mainline. Figure 1 depicts all types of risks using either unit or manifest trains transporting hazmat.



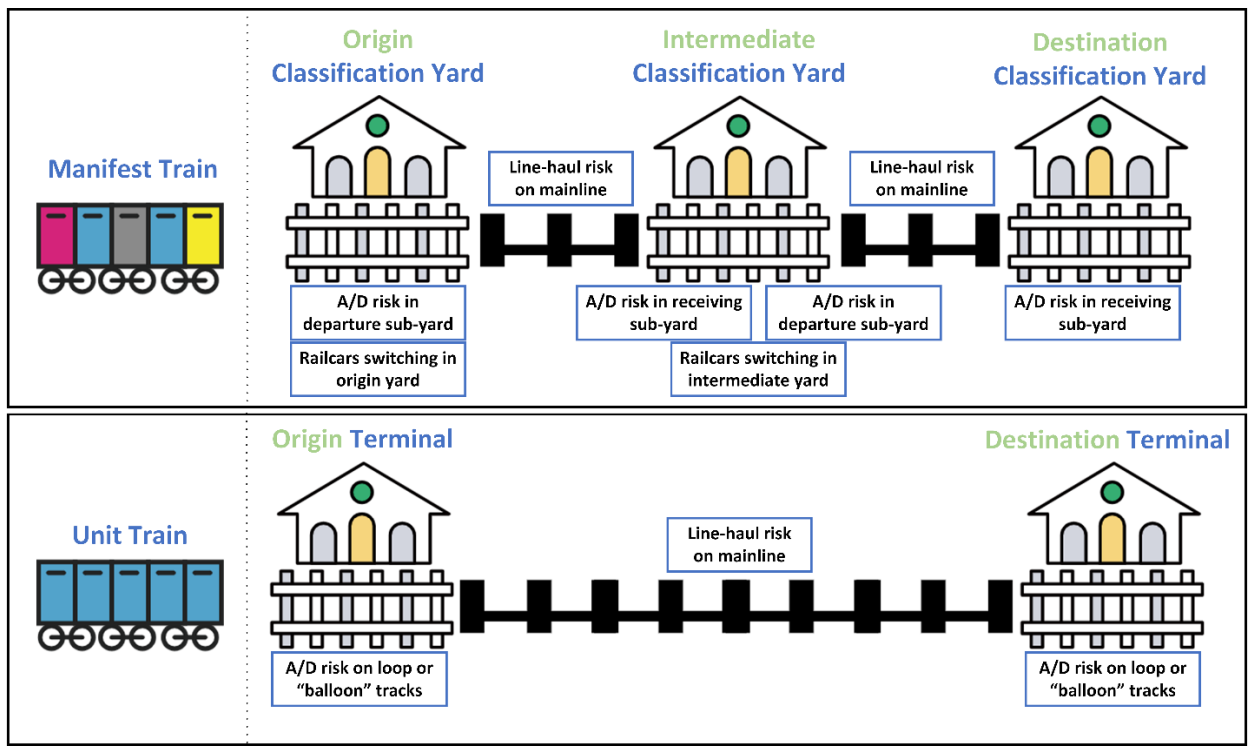

**Figure 1 Types of risks encountered by different train configurations.**

It is difficult to determine which service option experiences less risks than the other judging only based on train characteristics. The unit train provides economies of scale and short lead times, and it saves time and money by avoiding the complicated assembling and dissembling processes that pose additional risk in intermediate classification yards for manifest trains. However, since unit trains usually have more railcars on one shipment than manifest trains, unit train derailments result in more cars derailed per accident compared with manifest trains. Compared to unit train configurations, the placement of hazmat railcars in manifest trains, the switching approach, and the number and type of intermediate classification yards can affect hazmat transportation risks related to derailments and subsequent hazmat release for manifest trains. Previous work has explored transportation risks related to mainline and yard/terminal separately to some extent (Liu, 2017a; Zhao & Dick, 2022), factors that influence mainline and yard risks (Barkan et al., 2003),



and accident rates distinguishing mainline and yard (Anderson & Barkan, 2004). However, there is no comprehensive methodology that focuses on comparing the transportation risks between unit trains and manifest trains, with consideration to all mainline, yard, and terminal risk components.

By building the event-chain-based probabilistic risk analysis using multiple probabilistic models and the latest derailment rates analyzed based on data from 1996 to 2018 considering both mainline and yard/terminal operations, this paper proposes a comprehensive methodology comparing unit and manifest trains transporting a given amount of hazmat to answer the following question: given certain combinations of train lengths, tank car block size, operating speed, tank car placement in manifest trains, yard type, yard switching approach, and tank car design, will one train configuration experience less risk compared with the other?

The remainder of this paper is organized as follows: in this paper, Section 2 discusses previous work related to risk analyses of hazardous material trains. Section 3 presents the proposed methodology in detail, which considers both mainline and yard components. Section 4 concludes the whole paper.

## 2. Literature Review

Railroads in North America transported more than two million carloads of hazardous materials in 2018 (Bureau of Explosives, 2019). Although more than 99 percent of shipments reached their destinations safely and transportation of hazardous materials accounts for only 8.1 percent of rail traffic in the United States, this traffic is responsible for a major share of railroad liability and



insurance risk (Association of American Railroads, 2021). Over the past 20 years, railroads have decreased hazmat accident rates by 64 percent by increasing the safety design of tank cars carrying hazmat, choosing the safest routes to transport hazmat, and training first responders. However, a comprehensive risk analysis considering all possible risks a hazmat shipment might experience on both mainline and in yards/terminals is still lacking.

Existing studies have performed risk analyses of rail transport of hazardous material *on mainlines* based on the event chain. They have more or less explored the event chain towards the accident, and the risks have been quantified or modeled using different methodologies and historical datasets. To analyze the consequences of a train derailment involving the release of hazmat, the following elements have been studied in previous papers: the likelihood of a train derailment on a track segment (Dindar et al., 2018; Kaeeni et al., 2018; Lin et al., 2020; Nayak et al., 1983), the number of railcars derailed (Anderson, 2005; Bagheri, 2010; Martey & Attoh-Okine, 2019; Wang et al., 2020), the number of tank cars derailed (Bagheri et al., 2014; Kawprasert & Barkan, 2010; Saccomanno & El-Hage, 1991; Verma, 2011), the number of hazmat cars releasing contents (Barkan, 2008; Kawprasert & Barkan, 2010; Liu et al., 2014; Saat & Barkan, 2011), and the consequences (Branscomb et al., 2010; Glickman et al., 2007; Verma & Verter, 2007; Yoon et al., 2009). Bagheri et al. (2014) used the truncated geometric model to estimate the number of cars derailing and then multiplied the conditional probability of release to estimate the number of hazmat cars derailing and releasing. However, they did not consider different train configurations (especially manifest trains) or the probability distribution of release quantity. Liu (2017a) improved the model that Bagheri et al. (2014) developed by considering different derailing



probabilities at different train positions. They focused on probability analysis of a release incident without accounting for the number of hazmat cars releasing contents or the consequences.

Typically, manifest trains carrying hazardous material shipments pass through numerous yards and are switched between trains several times during each trip. Despite the amount of time spent in classification yards, these facilities tend to be deemphasized or excluded in railway hazardous materials transportation risk assessments (Center for Chemical Process, 1995). Of the three aspects of the rail transportation process, the risk of the line-haul movement along mainline routes, intermediate yards, and loading/unloading captures the majority of attention when evaluating risk (Purdy et al., 1988). Glickman & Erkut (2007) argued that while the risk of movement through rail yards cannot be ignored, yard risk receives little attention due to the perception of it being a minor risk compared to the mainline and a lack of data to support analysis.

The rapid increase in crude oil transportation by rail in North America in recent years has led to more hazardous material being moved in dedicated unit trains of 80 to 100 railcars that transport a single commodity in a continuous movement from origin to destination (Dick & Brown, 2014). Moving large amounts of hazardous material in a single unit train may increase the risk of multiple-car releases and the disproportionately large consequences of large-release events (Liu, 2017a; Liu et al., 2014). However, these unit trains bypass intermediate classification yards and thus avoid the risk associated with sorting railcars in these facilities. On the other hand, if railcars are placed in the lowest-probability derailment positions as suggested in previous research (Bagheri, 2010; Bagheri et al., 2011), distributing them to multiple manifest trains would result in a lower risk level



(Liu, 2017b). Fully quantifying this risk trade-off requires a better understanding of the risk of handling hazardous material railcars both on mainlines and in classification yards/terminals.

Building a comprehensive risk assessment methodology considering potential risks on mainlines and in yards/terminals is also important for analyzing strategies, such as railcar placement, that are intended to reduce the line-haul risk of railway hazardous materials transportation (Liu et al., 2013). Railcar placement reduces risk by moving railcars carrying hazardous materials to the portion of the train with the lowest probability of release (Federal Railroad Administration, 2005). Achieving optimal car placement for safety requires deviating from conventional operating practices in classification yards that minimize the amount of railcar handling, thus introducing risk associated with additional car coupling events (Branscomb et al., 2010). Increased switching activity also increases the risk of railway employee injury (An et al., 2007). Previous studies of this approach have either ignored the risk in classification yards while focusing on the line-haul route risk or have only acknowledged a potential increase in risk associated with the classification yard train assembly process in a cursory manner as a subject for future research (Bagheri, 2010; Bagheri et al., 2011, 2012; Federal Railroad Administration, 2005).

Barkan et al. (2003) calculated railcar derailment rates for both mainlines and yards but concentrated on mainline accidents for subsequent research. Yard accidents were deemphasized since the yard incidents occurred at low speeds and were less likely to lead to hazardous material release (Anderson & Barkan, 2004). In making the decision to focus on mainline train accidents, since the risk of railroad hazardous materials transport is the product of the likelihood of a release event and its consequence (Saat et al., 2014), Barkan et al. (2003) only considered the likelihood



half of the risk equation. They did not fully consider the consequence of population exposure, implicitly assuming that it would be identical along mainline routes and surrounding yards and terminals. However, while many mainline route-miles are in remote, sparsely populated rural areas, most classification yards are located in urban areas that are moderately to densely populated, increasing human exposure to potential releases (Christou, 1999).

With less focus on railroad yard accident risk, there has been a lack of research analyzing risks throughout the shipment which considers both mainline and yard/terminal components. This two-part paper is the first study to build event chains for all line-haul risks, arrival/departure risks, and yard switching risks while considering train configuration. In this paper (Part I), yard types, train types, yard switching approaches, and a series of probability distributions for each event chain component are discussed in detail when building the probabilistic models. Instead of ignoring the potential risks caused by arrival/departure events and yard switching events as previous work did, this paper quantifies potential risks during these two events together with line-haul risks on mainlines. Additionally, the event chain for mainline risk analysis is also extended, modified, and improved. Risks are modeled as the total expected casualties considering route characterization, weather characterization, and evacuation response time. By considering risks on mainlines and yards/terminals for unit trains and manifest trains, this paper provides a solid approach for any train configuration transporting any amount of hazmat on any planned railroad. The proposed methodology can be used as a calculator to guide the operational arrangement in practice to reduce the total potential risks encountered.



### 3. Probabilistic Risk Analysis Methodology

The fundamental operating differences on mainlines and in yards/terminals determine the different components required to model the transportation risks by rail. Figure 2 shows the event chains for three different types of risks: line-haul risks on mainlines, A/D risks in yards (for manifest trains) or terminals (for unit trains), and yard switching risks (for manifest trains). In this section, three event chains for these three types of events will be elaborated component-by-component in the following subsections. All notations used in this paper are summarized in

Table 1.

**Table 1 Notations for mainline risks and yard/terminal risks**

***Notations related to risk components on mainlines***

| | |
|---|---|
| $P_{i,c\|c\in\text{TRM}}$ | The probability that the derailment happens on mainline segment $i$ and it is caused by a mainline train-mile-based cause. |
| $P_{i,c\|c\in\text{TOM}}$ | The probability that the derailment happens on mainline segment $i$ and it is caused by a mainline ton-mile-based cause. |
| $P_{i,c\|c\in\text{CM}}$ | The probability that the derailment happens on mainline segment $i$ and it is caused by a mainline car-mile-based cause. |
| $d_{\text{TRM}}$ | The number of mainline train derailments per million mainline train-miles. |
| $d_{\text{TOM}}$ | The number of mainline train derailments per billion mainline gross ton-miles. |
| $d_{\text{CM}}$ | The number of mainline train derailments per billion mainline car-miles. |
| $p_c$ | Percentage of mainline train derailments due to $c^{\text{th}}$ cause in the total number of mainline derailments. |
| $L$ | Train length, i.e., number of railcars in the train. |
| $L_i$ | The length (in miles) of the track segment $i$ where the train will travel from the origin to the destination. |
| $GW$ | Gross tonnage of the train (including lightweight and lading). |
| $PTD_{i,\text{main}}$ | The probability of a train derailment on mainline segment $i$. |
| $TD$ | A train derailment, which can be a line-haul train derailment. |
| $POD(k\|TD)$ | The conditional probability that the POD is at the $k^{\text{th}}$ position in a train given a train derailment (which can be a line-haul incident). |
| $POD$ | The point of derailment. |
| $F(x)$ | The cumulative density distribution of the corresponding fitted NPOD distribution (i.e., best fitted Beta distributions). |
| $P_i(x\|POD= k)$ | The conditional probability of derailing $x$ cars given that the POD is at $k^{\text{th}}$ position on segment $i$. |
| $L_r$ | The number of railcars behind the point of derailment. |
| $GT$ | The average gross tonnage per car. |
| $EUT$ | If the train is an empty unit train, $EUT = 1$, otherwise $EUT = 0$. |
| $LUT$ | If the train is a loaded unit train, $LUT = 1$, otherwise $LUT = 0$. |
| $PD_i(j\|TD)$ | The conditional probability of derailing the car at $j^{\text{th}}$ position on track segment $i$ given a train derailment. |
| $R_i(j\|TD)$ | The conditional probability of releasing the car at the $j^{\text{th}}$ position in a train on segment $i$ given a train derailment. |
| $CPR$ | The base conditional probability of release for a tank car with a specific tank car type developed in Treichel et al. (2019). |
| $I_t(j)$ | The 0-1 indicator, equals 1 if the car at the $j^{\text{th}}$ position in a train is a tank car, and 0 otherwise. |
| $y_i$ | The occurrence of a train derailment at position $j$ in a train. |
| $P_{\text{main,i,re}}(x_R\|TD)$ | The conditional probability of releasing $x_R$ tank cars on the mainline segment $i$ given a train derailment. |
| $P_{\text{main,i,re}}(x_R)$ | The probability of releasing $x_R$ tank cars on the mainline segment $i$ per shipment. |
| $P_{re}(x)$ | The probability of releasing $x$ gallons of contents in total from all releasing tank cars (from Section 3.5). |
| $C(x,t)$ | The expected total casualties caused by releasing $x$ gallons of content at response time $t$, and $t \in [0,120]$ in minutes (Figure 5). |
| $TC(t)$ | The expected total casualties after $t$ minutes caused by a releasing incident. |
| $\delta$ | The number of tank cars that need to be transported. |
| $c_{\text{unit}}$ | The capacity of the unit train, i.e., the number of tank cars a unit train can carry. |
| $c_{\text{manifest}}$ | The capacity of the manifest train, i.e., the number of tank cars a manifest train can carry. |

***Notations related to risk components in yard/terminal***



| | |
|---|---|
| $P_{c\|c\in ADTR}$ | The probability that this freight consist train derails while arriving at or departing from a terminal or classification yard and the derailment is caused by a train-processed-based cause. |
| $P_{c\|c\in ADCA}$ | The probability that this freight consist train derails while arriving at or departing from a terminal or classification yard and the derailment is caused by a car-processed-based cause. |
| $P_{YS}$ | The probability that this yard consist train derails while switching in yards. |
| $d_{ADTR}$ | The number of A/D train derailments per million train A/D events. |
| $d_{ADCA}$ | The number of A/D train derailments per billion car A/D events. |
| $d_{YS}$ | The number of yard switching derailments per million cars processed in the yard. |
| $p_c$ | Percentage of derailments of $c^{th}$ cause in the total number of derailments while arriving at or departing from a terminal or classification yard. |
| $L$ | Train length, i.e., number of railcars in the train. |
| $n$ | The number of arrival/departure events that a shipment involves. |
| $m$ | The number of intermediate yards that a manifest train shipment involves. |
| $PTD_{ADI}$ | The probability of a train derailment per shipment during A/D events. |
| $PTD_{YSI}$ | The probability of a train derailment per shipment during yard switching events. |
| $TD$ | A train derailment, which can be an A/D train derailment. |
| $POD(k\|TD)$ | The probability that POD is at the $k^{th}$ position in a train given a train accident (which can be an A/D derailment). |
| $POD$ | The point of derailment. |
| $F(x)$ | The cumulative density distribution of the corresponding fitted NPOD distribution (i.e., best fitted Beta distributions). |
| $P_i(x\|POD=k)$ | The conditional probability of derailing $x$ cars given that the POD is at the $k^{th}$ position in a train on segment $i$ per A/D incidents. |
| $f(x)$ | The probability mass functions of the best fitted generalized exponential distributions estimating the number of railcars derailed given the first car of derailment. |
| $YSI$ | A yard switching incident. |
| $P_{YardDeRail}(x\|FCD=k)$ | The conditional probability of derailing $x$ railcars given that the first car of the derailment is at the $k^{th}$ position in the group of cars. |
| $PD_i(j)$ | The conditional probability of derailing the car at $j^{\text{th}}$ position on track segment $i$ given a derailment. |
| $ADI$ | An arrival/departure incident in the yard/terminal. |
| $P_{\text{A/DDe}}(x_{\text{tank}}\|ADI)$ | The conditional probability of derailing $x_{\text{tank}}$ tank cars given an A/D incident. |
| $\delta_t(j)$ | 0-1 indicator, equals 1 if the car at the $j^{\text{th}}$ position in the train is a tank car, and 0 otherwise. |
| $P_{\text{A/DRe}}(x_{\text{tank}}\|ADI)$ | The conditional probability that there are $x_{\text{tank}}$ hazmat cars releasing contents given an A/D incident in yards/terminals. |
| $TT$ | The total number of tank cars in a yard switching event. |
| $CPR$ | The base conditional probability of release developed in Treichel et al. (2019). |
| $FCD(k\|YSI)$ | The probability that the first car of derailment is at the $k^{\text{th}}$ position in the group of cars switched together given a yard switching incident (for yard switching incidents only). |
| $FCD$ | The position of the first car of derailment in the group of cars switched together. |
| $TCC$ | The total number of cars considered in a yard switching event. For "switched alone" approach, $TCC$ is the number of tank cars, while it is the number of tank cars plus 19 non-tank cars for the "switched en masse" approach. |
| $P_{\text{YardDeTank}}(x_{\text{tank}}\|YSI)$ | The conditional probability of derailing $x_{\text{tank}}$ tank cars given a yard switching incident. |
| $P_{\text{terminal}}(x_{\text{tank}})$ | The probability of releasing $x_{\text{tank}}$ tank cars per shipment for a unit train in terminals. |
| $P_{\text{yard}}(x_{\text{tank}})$ | The probability of releasing $x_{\text{tank}}$ tank cars per shipment for a manifest train in yards. |
| $PTD_{\text{ADI,Unit}}$ | The probability of a train derailment per shipment during A/D events in terminals using unit trains. |
| $PTD_{\text{ADI,Manifest}}$ | The probability of a train derailment per shipment during A/D events in the yard using manifest trains. |
| $PTD_{\text{SWI}}$ | The probability of a train derailment per shipment during yard switching events. |
| $P_{\text{YardReTank}}(x_{\text{tank}}\|YSI)$ | The conditional probability that there are $x_{\text{tank}}$ hazmat cars releasing contents given a yard switching incident. |
| $P_{\text{re}}(x)$ | The probability of releasing $x$ gallons of contents in total from all releasing tank cars (from Section 3.5). |
| $C(x,t)$ | The expected total casualties caused by releasing $x$ gallons of content at response time $t$, and $t \in [0,120]$ in minutes (Figure 5). |
| $TC(t)$ | The expected total casualties after $t$ minutes caused by a releasing incident. |
| $\delta$ | The number of tank cars that need to be transported. |
| $c_{\text{unit}}$ | The capacity of the unit train, i.e., the number of tank cars a unit train can carry. |
| $c_{\text{manifest}}$ | The capacity of the manifest train, i.e., the number of tank cars a manifest train can carry. |



| | Mainline incident | Yard or terminal: arrival/departure incident | Yard: switching incident |
|---|---|---|---|
| **Step 1** | **Probability of a line-haul incident on the mainline (Section 3.1.1)**<br>- Derailment causes<br>- Train configuration | **Probability of an A/D incident in the yard or terminal (Section 3.1.2)**<br>- Derailment causes<br>- Train configuration<br>- Yard type<br>- Number of intermediate yards | **Probability of a yard switching incident (Section 3.1.2)**<br>- Derailment causes<br>- Train length<br>- Number of intermediate yards |
| **Step 2** | **Probability of derailing certain number of cars (Section 3.2.1)**<br>- Train length<br>- Point of derailment<br>- Average gross tonnage<br>- Derailment speed<br>- Train configuration | **Probability of derailing certain number of cars (Section 3.2.1)**<br>- Point of derailment<br>- Train length | **Probability of derailing certain number of cars (Section 3.2.2)**<br>- Train length<br>- Yard type<br>- **The position of the first car of derailment in the group of cars switched together** |
| **Step 3** | **Probability of derailing at certain position (Section 3.3.1)**<br>- Point of derailment<br>- Derailment severity given the point of derailment | **Probability of derailing certain number of tank cars (Section 3.3.2)**<br>- Point of derailment<br>- Position-dependent derailment probability<br>- Train length<br>- Train configuration<br>- Placement of tank cars | **Probability of derailing certain number of tank cars (Section 3.3.3)**<br>- Switching approach<br>- Number of tank cars<br>- Number of non-tank cars switched together<br>- The position of the first car of the derailment in the group of cars switched together |
| **Step 4** | **Probability of releasing at certain position (Section 3.3.1)**<br>- Conditional probability of release<br>- Tank car type<br>- Derailment speed<br>- Probability of derailment at certain position | **Probability of releasing certain number of tank cars (Section 3.3.2)**<br>- Conditional probability of release<br>- Probability distribution of the number of tank cars derailed | **Probability of releasing certain number of tank cars (Section 3.3.3)**<br>- Conditional probability of release<br>- Probability distribution of the number of tank cars derailed |
| **Step 5** | **Probability of releasing certain number of tank cars (Section 3.3.1)**<br>- Placement of tank cars in a train<br>- Probability of releasing at certain position | **Probability of releasing certain amount of content (Section 3.5)**<br>- Tank car safety design<br>- Derailment speed<br>- Probability distribution of the number of tank cars releasing content | **Probability of releasing certain amount of content (Section 3.5)**<br>- Tank car safety design<br>- Derailment speed<br>- Probability distribution of the number of tank cars releasing content |
| **Step 6** | **Probability of releasing certain amount of content (Section 3.5)**<br>- Tank car safety design<br>- Derailment speed<br>- Probability distribution of the number of tank cars releasing content | **Consequence (Section 3.6)**<br>- Route characterization<br>- Weather characterization<br>- Response time | **Consequence (Section 3.6)**<br>- Route characterization<br>- Weather characterization<br>- Response time |
| **Step 7** | **Consequence (Section 3.6)**<br>- Route characterization<br>- Weather characterization<br>- Response time | | |

**Figure 2 Flow chart of event-chain-based risk analysis and the related risk components.**



### 3.1 Train Derailment Probability

### 3.1.1 Derailments on Mainlines

According to Liu (2015), when traffic exposure is large and the derailment rate (i.e., the number of derailments normalized by the corresponding traffic volume) is relatively low, the probability of train derailment can be numerically approximated by multiplying the derailment rate by the mileage of the train shipment. Thus, the probability of train derailment can be estimated based on the train derailment rate using historical train derailment data and traffic data. The FRA has categorized more than 300 accident causes into five groups based on the circumstances and conditions of accidents (FRA, 2012). The hierarchically organized groups can be classified as track, equipment, human factors, signal, and miscellaneous, with each cause being assigned a unique cause code. In the 1990s, a study by Arthur D. Little, Inc (ADL) grouped similar FRA accident causes together based on experts' opinions, producing a variation on the FRA subgroups (Arthur D. Little, Inc. (ADL), 1996). Previous studies (Liu, 2015, 2016; Liu et al., 2012; Schafer & Barkan, 2008) found that the ADL cause groups could be more fine-grained, allowing for greater resolution for certain causes. For example, the FRA combines broken rails, joint bars, and rail anchors in the same subgroup, whereas the ADL grouping distinguishes between broken rail and joint bar defects. Thus, this study uses ADL cause groups to conduct its cause-specific railroad derailment analysis.

The traffic volume data used in this paper is obtained from Class I railroads[1] annual report financial data and Surface Transportation Board (STB) waybill sample data, which is available for the years between 1996 and 2018 at the time of this analysis (Dick et al., 2021). We count the number of accidents that occurred by their cause category. In total, there were 2,462 unit train derailments and 5,514 manifest train derailments on mainlines over these years. These accidents are classified into 46 cause groups. Table A.1 in Appendix A shows train derailment data from 1996 to 2018 by cause group and train type on Class I mainlines.

---

[1] As of 2019, the Surface Transportation Board defines a Class I as having operating revenues of, or exceeding, $505 million annually. (Resource: https://en.wikipedia.org/wiki/Railroad_classes.)

This study develops a cause-based train-derailment probability model. First, train derailment causes are classified into three categories: train miles, ton miles, and car miles (railcars only), respectively. For example, "broken wheels" could be associated with car miles traveled, and thus the probability of derailment caused by "broken wheels" should be calculated based on the traffic metric "car miles." In contrast, obstruction-caused accidents may be affected by the number of trains, and thus the probability of derailment caused by "obstruction" could be calculated based on the traffic metric "train miles."

Let $TRM$ denote the set of train-mile-based derailment causes, $TOM$ be the set of ton-mile-based derailment causes, and $CM$ be the set of car-mile-based derailment causes. If a train has $L$ railcars, its gross tonnage is denoted as $GW$, and it travels on a track segment $i$ with length of $L_i$ (in miles). The probability of train derailment due to the $c^{\text{th}}$ cause can be calculated by:

$$P_{i,c|c \in \text{TRM}} \approx d_{\text{TRM}}/1{,}000{,}000 \times L_i \times p_c \tag{1}$$

$$P_{i,c|c \in \text{TOM}} \approx d_{\text{TOM}}/1{,}000{,}000{,}000 \times GW \times L_i \times p_c \tag{2}$$

$$P_{i,c|c \in \text{CM}} \approx d_{\text{CM}}/1{,}000{,}000{,}000 \times L \times L_i \times p_c \tag{3}$$

where



$P_{i,c|c\in\text{TRM}}$: the probability that the derailment happens on mainline segment $i$ and it is caused by a mainline train-mile-based cause.

$P_{i,c|c\in\text{TOM}}$: the probability that the derailment happens on mainline segment $i$ and it is caused by a mainline ton-mile-based cause.

$P_{i,c|c\in\text{CM}}$: the probability that the derailment happens on mainline segment $i$ and it is caused by a mainline car-mile-based cause.

$d_{\text{TRM}}$: the number of mainline train derailments per million mainline train-miles (Table 2).

$d_{\text{TOM}}$: the number of mainline train derailments per billion mainline gross ton-miles (Table 2).

$d_{\text{CM}}$: the number of mainline train derailments per billion mainline car-miles (Table 2).

$p_c$: percentage of mainline train derailments due to $c^{\text{th}}$ cause in the total number of mainline derailments (Table A.1 in Appendix A).

$L$: train length, i.e., number of railcars in the train.

$L_i$: the length (in miles) of the track segment $i$ where the train will travel from the origin to the destination.

$GW$: gross tonnage of the train (including lightweight and lading).

The derailment rate by traffic metric data shown in Table 2 is calculated by Zhang et al. (2022) using FRA-reportable Class I mainline train derailment data for the years 1996-2018. Since the train derailment probability per train shipment is very minimal, the probability of a train derailment on mainline segment $i$ (denoted as $PTD_{i,\text{main}}$) can be estimated by:

$$PTD_{i,\text{main}} \approx \sum_{c\in\text{TRM}} P_{i,c|c\in\text{TRM}} + \sum_{c\in\text{TOM}} P_{i,c|c\in\text{TOM}} + \sum_{c\in\text{CM}} P_{i,c|c\in\text{CM}} \qquad (4)$$



**Table 2 Derailment rate on mainlines by traffic metric (Zhang et al., 2022)**

**(a) Unit train**

| Metric | Derailments |
| --- | --- |
| Derailments per million train-miles | 0.85 |
| Derailments per billion gross ton-miles | 0.10 |
| Derailments per billion car-miles | 8.14 |

**(b) Manifest train**

| Metric | Derailments |
| --- | --- |
| Derailments per million train-miles | 0.67 |
| Derailments per billion gross ton-miles | 0.14 |
| Derailments per billion car-miles | 11.39 |

### 3.1.2 Derailments in Yards and Terminals

There are two types of events that can cause a derailment in the yard and terminal: the *arrival/departure event* (A/D event) for unit trains (or manifest trains) arriving at or departing from terminal facilities (or classification yards), and the *yard switching event* associated with the sorting, switching, assembling, and dissembling processes for manifest trains in classification yards. According to Zhao and Dick (2022), the A/D events are classified as either train-mile-based causes or car-mile-based causes. Thus, the probability of an A/D event is estimated by the cause-based train derailment model. Assume there is a manifest train with $L$ cars (railcars only) and the



manifest train transverses $m$ intermediate classification yards with $n$ A/D events. The relationship between $n$ and $m$ for manifest trains is developed in Equation (5). Since each railcar is switched once at the origin yard and once at each intermediate yard, this manifest train has $(m + 1) \times L$ car switching movements. Similarly, by definition, a unit train with $L$ railcars (railcars only) will have two A/D events (one at the origin yard and one at the destination yard). On the other hand, the yard switching derailment depends on the number of cars possessed.

The probability of an A/D train derailment due to the $c^{\text{th}}$ cause can be calculated by Equation (6) if $c$ is a train-mile-based cause and by Equation (7) if $c$ is a car-processed-based cause. If the train derailment is due to a yard switching event (for manifest trains only), Equation (8) is used to calculate the probability of a train derailment.

$$n = 1 \text{ (origin yard)} + 2 \times m \text{ (intermediate yard)} +$$
$$1 \text{ (destination yard)} \tag{5}$$

$$P_{c|c \in \text{ADTR}} \approx d_{\text{ADTR}}/1{,}000{,}000 \times n \times p_c \tag{6}$$

$$P_{c|c \in \text{ADCA}} \approx d_{\text{ADCA}}/1{,}000{,}000{,}000 \times L \times n \times p_c \tag{7}$$

$$P_{\text{YS}} \approx d_{\text{YS}}/1{,}000{,}000 \times L \times (m + 1) \tag{8}$$

$P_{c|c \in \text{ADTR}}$: the probability that this freight consist train derails while arriving at or departing from a terminal or classification yard and the derailment is caused by a train-processed-based cause.

$P_{c|c \in \text{ADCA}}$: the probability that this freight consist train derails while arriving at or departing from a terminal or classification yard and the derailment is caused by a car-processed-based cause.

$P_{\text{YS}}$: the probability that this yard consist train derails while switching in yards.

$d_{\text{ADTR}}$: the number of A/D train derailments per million train A/D events (Table 3).

$d_{\text{ADCA}}$: the number of A/D train derailments per billion car A/D events (Table 3).



$d_{\text{YS}}$: the number of yard switching derailments per million cars processed in the yard (Table 3).

$p_c$: percentage of derailments of $c^{\text{th}}$ cause in the total number of derailments while arriving at or departing from a terminal or classification yard (Table B.1 in Appendix B).

$L$: train length, i.e., number of railcars in the train.

$n$: the number of arrival/departure events that a shipment involves.

$m$: the number of intermediate yards that a manifest train shipment involves.

**Table 3 Derailment rates for various events, train configurations, yard types, and yard traffic metrics for years 1996-2018 (Zhao & Dick, 2022)**

| | Arrival/Departure event | | Yard switching event |
|---|---|---|---|
| *Group* | *Derailments per million train arrival/departures* | *Derailments per million car arrival/departures* | *Derailments per million cars-processed in classification yards* |
| Manifest train | 61.52 | 1.04 | 6.43 |
| Flat yard | 118.92 | 2.02 | 6.38 |
| Hump yard | 36.53 | 0.62 | 6.49 |
| Unit train | 76.95 | 0.74 | N/A |
| Loaded unit | 126.31 | 1.22 | N/A |

Combining risk components due to various causes, the probability of a train derailment per shipment during A/D events and during yard switching events, defined as $PTD_{\text{ADI}}$ and $PTD_{\text{YSI}}$ can be approximately estimated as follows:



$$PTD_{\text{ADI}} \approx \sum_{c \in \text{ADTR}} P_{c|c \in \text{ADTR}} + \sum_{c \in \text{ADCA}} P_{c|c \in \text{ADCA}} \qquad (9)$$

$$PTD_{\text{YSI}} \approx P_{\text{YS}} \qquad (10)$$

Note that the calculation of $PTD_{\text{ADI}}$ and $PTD_{\text{YSI}}$ can distinguish train types, yard types, and yard switching approaches by considering different derailment datasets.

### 3.2 Number of Railcars Derailed Per Train Derailment

#### 3.2.1 Line-haul Incidents on Mainlines and A/D Incident in Yards/Terminals

Derailment severity, defined as the total number of railcars derailed given a mainline train derailment, can be affected by the point of derailment (POD), derailment speed, train type, train length (number of railcars), and average gross tonnage per car on mainlines. Since the arrival/departure process in a yard or terminal operates similarly to mainline freight operation with a reduced speed, the method to estimate the derailment severity of an A/D accident in yard and terminal is the same as it is on mainlines.

Normalized by the train length (the number of railcars), the normalized POD (denoted as NPOD) can be best predicted by Beta distributions of $Beta(0.7549, 0.9582)$ for the unit train and $Beta(0.7842, 1.1002)$ for the manifest train on mainlines using FRA train derailment data from 1996 to 2018. The "best fits" are $Beta(0.5350, 0.9121)$ and $Beta(0.7729, 0.9034)$ for the manifest train and the unit train in yard and terminal. The Beta distribution fits are consistent with



findings from prior research using older datasets (Saccomanno & El-Hage, 1989, Saccomanno & El-Hage, 1991, and Liu et al., 2014).

The probability that the train derails starting from the $k^{\text{th}}$ position (for both mainline and A/D derailments) can be calculated by (Liu et al., 2014; Liu & Schlake, 2016):

$$POD(k|TD) = F\left(\frac{k}{L}\right) - F\left(\frac{k-1}{L}\right) \qquad (11)$$

$TD$: a train derailment, which can be a line-haul train derailment or an A/D train derailment.

$POD(k|TD)$: the probability that the POD is at the $k^{\text{th}}$ position of a train given a train derailment.

$F(x)$: the cumulative density distribution of the corresponding fitted NPOD distribution (i.e., best fitted Beta distributions).

$L$: train length, i.e., number of railcars in the train.

As demonstrated by previous studies (Anderson & Barkan, 2005; Bagheri et al., 2011; Saccomanno & El-Hage, 1989, 1991), the conditional probability of derailing $x$ railcars given that the point of derailment is at the $k^{\text{th}}$ position on segment $i$, denoted as $P_i(x|\text{POD}= k)$ can be estimated by the Truncated Geometric Logistic model:

$$P_i(x|\text{POD}= k) = \begin{cases} \frac{\frac{\exp(z)}{1+\exp(z)} \times \frac{\exp(z)}{[1+\exp(z)]^{x-1}}}{1-\left[\frac{1}{1+\exp(z)}\right]^{L_r}}, & \text{if } x = 1,2,\dots L_r \\ 0, & \text{otherwise} \end{cases} \qquad (12)$$

$$L_r = L - POD + 1 \qquad (13)$$

where $z$ takes different values for different derailment locations:



$$z_{\text{main}} = -0.952 - 0.0306 \times DS - 0.0018 \times L_r - 0.00239 \times GT \qquad (14)$$

$$+ 0.119 \times EUT - 0.339 \times LUT$$

$$z_{\text{yard}} = -1.595 - 0.0029 \times L \qquad (15)$$

$$z_{\text{term}} = -1.574 - 0.0016 \times L \qquad (16)$$

where

$P_i(x|\text{POD}= k)$: the conditional probability of derailing $x$ railcars given that the POD is at the $k^{\text{th}}$ position on segment $i$.

$L$: train length, i.e., number of railcars in the train.

$POD$: point of derailment.

$L_r$: the number of railcars behind the point of derailment, defined in Equation (13).

$GT$: average gross tonnage per car.

$EUT$: if the train is an empty unit train, $EUT = 1$, otherwise $EUT = 0$.

$LUT$: if the train is a loaded unit train, $LUT = 1$, otherwise $LUT = 0$.

On mainlines, the derailment severity can be estimated using Equations (12) - (14). The parameters used in Equation (14) are from Liu et al. (2022). When building this model, the manifest train is used as a reference. Thus, for manifest trains, variables $EUT$ and $LUT$ in Equation (14) are set to 0. In the yard, the derailment severity for a manifest train A/D incident is estimated by Equations (12), (13), and (15); in the terminal, the derailment severity for a loaded unit train A/D incident can be calculated by Equations (12), (13), and (16) (Liu et al., 2022).



### 3.2.2 Yard Switching Events

While the study of arrival/departure risk in the yard/terminal can consider the same unit train and manifest train consists studied on the mainline, by definition, the yard switching process will alter the arriving manifest train consist into new departing manifest train consists to the same destination yard. The yard switching process typically involves the movement of a single railcar, a cut of cars, or a portion of a train (potentially moving in reverse or as a shoving movement) at a reduced speed by a yard switching crew using a switch engine (not the mainline locomotive). Thus, the traditional definitions of a train consist and "point of derailment" described in Section 3.2.1 are not applicable, and a new risk analysis methodology for yard switching incidents should be developed.

Liu et al. (2022) examined 89 potential distribution models and found that the generalized exponential distribution best fits the empirical FRA Rail Equipment Accident (REA) yard derailment data for the years 1996-2018. The probability mass functions (denoted as $f(x)$) of the best fitted generalized exponential distributions for yard switching events are presented in Equation (17) for all yard types, in Equation (18) for flat yards, and in Equation (19) for hump yards.

$$f(x) = \left(1.44 + 1.37^{-7} \times (1 - e^{-1.1x})\right) \times \exp^{1.44x - 1.37^{-7}x + 1.25^{-7} \times (1 - e^{-1.1x})} \qquad (17)$$

$$f(x) = \left(1.01 + 1.68^{-7} \times (1 - e^{-1.68x})\right) \times \exp^{1.01x - 1.68^{-7}x + 1.00^{-7} \times (1 - e^{-1.68x})} \qquad (18)$$

$$f(x) = (5.05^{-8} \qquad (19)$$
$$+ 2.40^{-5} \times (1 - e^{-3.12x}))$$
$$\times \exp^{5.05^{-8}x - 2.40^{-5}x + 7.70^{-6} \times (1 - e^{-3.12x})}$$



This paper assumes that the derailment occurs when a cut of the group of railcars are switched together for yard switching incidents. The concept of "point of derailment" in Section 3.2.1 is defined from the perspective of root causes of the train derailment (for example, the derailment frequently happens from the head to the end of a train), while the "first car of derailment (FCD)" for yard switching incidents is defined only as a "label" referring to the first vehicle derailed in the cut of derailed vehicles. The main difference between POD and FCD is that POD is defined for a regular freight train consist, while FCD is defined for a yard switch train consist. These two terms are defined separately to emphasize that the train consist during yard switching events is not the same as that on mainlines.

Empirical data indicates that manifest trains infrequently derail more than 20 cars in a yard switching incident. Thus, it is necessary to truncate the generalized exponential distribution to fit the empirical data, the length of the train (number of railcars), and the known first car of derailment. The conditional probability of derailing $x$ railcars in a yard switching incident given that the first car of the derailment is at the $k^{\text{th}}$ position in the group of cars, denoted as $P_{\text{YardDeRail}}(x|FCD = k)$, can be calculated by:

$$P_{\text{YardDeRail}}(x|FCD = k) \tag{20}$$

$$= \begin{cases} f(x), & \text{if } x < \min(20, L - k + 1) \\ \sum_{x=\min(20, \ L-k+1)}^{\infty} f(x), & \text{if } x = \min(20, L - k + 1) \end{cases}$$

where



$FCD$: the position of the first car of the derailment in the group of cars switched together.

$P_{\text{YardDeRail}}(x|FCD = k)$: the conditional probability of derailing $x$ railcars given that the first car of the derailment is at the $k^{\text{th}}$ position in the group of cars.

$f(x)$: the probability mass functions of the best fitted generalized exponential distributions defined in Equation (17) for all yard types, in Equation (18) for flat yards, and in Equation (19) for hump yards, respectively.

$L$: train length, i.e., number of railcars in the train.

### 3.3 Number of Hazmat Cars Releasing Contents Per Train Derailment

Different risk components (line-haul risks on mainlines, A/D risks in yards/terminals, and yard switching risks) follow different approaches to obtain the number of hazmat cars releasing contents. Before obtaining the probability distribution of the number of hazmat cars releasing contents, the analysis of line-haul risks on mainlines calculates the position-dependent releasing probability, while the analysis of A/D risks and the yard switching risks estimates the probability distribution of the number of hazmat cars derailed first (see Figure 2).

#### 3.3.1   Line-haul Events on Mainlines

Based on the calculated probability distribution of the total number of railcars derailed from Section 3.2.1, we can further calculate the conditional probability of the car at $j^{\text{th}}$ position derailing on track segment $i$ (defined as $PD_i(j|TD)$) given a train derailment on mainlines, which is the accordance to determine the train consist. Based on previous experience and historical data, it is assumed that if a train derailment occurs, cars will derail sequentially after the POD. For example,



if there are three vehicles derailed, they are POD, POD+1, and POD+2. According to previous work by Liu et al. (2018), $PD_i(j|TD)$ can be calculated by:

$$PD_i(j|TD) = \sum_{k=1}^{j} \left[ POD(k|TD) \times \sum_{x=j-k+1}^{L_r} P_i(x|POD=k) \right] \tag{21}$$

where

$TD$: a train derailment.

$PD_i(j|TD)$: the conditional probability of derailing the car at $j^{\text{th}}$ position on track segment $i$ given a train derailment.

$POD(k|TD)$: the probability that POD is at the $k^{th}$ position in a train given a train derailment.

$P_i(x|POD=k)$: the conditional probability of derailing $x$ railcars given that the POD is at the $k^{th}$ position in a train on segment $i$.

$\sum_{x=j-k+1}^{L_r} P_i(x|POD=k)$ is the sum of the probability that the locomotive or the railcar at the $j^{th}$ position is derailed, given that the POD is at $k^{\text{th}}$ position.

In the next step, the position-dependent derailment probability is extended to the position-dependent tank car releasing probability, given a train derailment. Let $I_t(j)$ be the 0-1 indicator, which equals 1 if the car at $j^{\text{th}}$ position of a train is a tank car, and 0 otherwise. We assume that the conditional probability of a derailed tank car releasing is the same given the same design and accident speed. It is also assumed that each tank car releases contents independently from others. These assumptions are made due to limited information regarding the relationship between the release probability of a derailed tank car and its position in a train. This paper calculates the



probability of release (CPR) for a tank car using the results included in the RSI-AAR Tank Car Safety Project (Treichel et al., 2019).

For a car at $j^{\text{th}}$ position of a train, the position-dependent tank car releasing probability on segment $i$ given a train derailment, which is denoted as $R_i(j|TD)$, can be calculated as:

$$R_i(j|TD) = PD_i(j|TD) \times [I_t(j) \times CPR] \qquad (22)$$

$R_i(j|TD)$: the conditional probability of releasing of the car at the $j^{\text{th}}$ position in a train on segment $i$ given a train derailment.

$PD_i(j|TD)$: the conditional probability of derailing the car at $j^{\text{th}}$ position on track segment $i$ given a train derailment.

$CPR$: the base conditional probability of release for a tank car with a specific tank car type developed in Treichel et al. (2019).

$I_t(j)$: the 0-1 indicator, equal to 1 if the car at the $j^{\text{th}}$ position in a train is a tank car, and 0 otherwise.

Based on the position-dependent tank car releasing probability given a train derailment, we can further calculate the probability distribution of the number of tank cars releasing contents. Let $y_j$ represent whether the tank car at $j^{\text{th}}$ position releases content, which is a 0-1 variable. For each car in a train, whether a tank car would release at $j^{\text{th}}$ position is a Bernoulli variable with releasing probability of $R_i(j|TD)$, and the probability of releasing could vary by position in a train (due to the position-dependent car derailment probability):

$$y_j \sim Bernoulli\big(R_i(j|TD)\big) \qquad (23)$$



For the entire train, the total number of tank cars releasing contents follows a Poisson Binomial distribution, which is the sum of independent Bernoulli random variables that are not necessarily identically distributed (Chen & Liu, 1997). The Poisson Binomial distribution is used to estimate the probability associated with a certain number of releasing tank cars in a group of derailed tank cars. Let $x_R$ be the total number of tank cars releasing contents and $L$ be the train length. For each tank car, whether it releases is a binary event (release or no release) with release probability $R_i(j|TD)$, $\forall j: y_j = 1$. $x_R$ can be mathematically expressed as Equation (24). It follows the Poisson Binomial distribution with mean of $\sum_{j=1}^{L} R_i(j|TD)$ and variance of $\sum_{j=1}^{L} R_i(j|TD) \times \left(1 - R_i(j|TD)\right)$.

$$x_R = \sum_{j=1}^{L} y_j \sim Poisson\ Binomial\ Distribution \tag{24}$$

### 3.3.2 Arrival/departure Events in Yards/Terminals

Section 3.2.1 calculates the probability that the point of derailment is at the $k^{\text{th}}$ position in a train and the probability of derailing $x$ railcars given the point of derailment is at the $k^{\text{th}}$ position in a train. The probability of derailing $x_{\text{tank}}$ tank cars given an A/D incident depends on train configuration and the placement of the block of tank cars in a manifest train, which can be calculated as:

$$P_{\text{A/DDe}}(x_{\text{tank}}|ADI)$$
$$= \sum_{k=1}^{L} \sum_{\forall x:\ x_{\text{tank}} = \sum_{j=k}^{k+x} \delta_t(j)} POD(k|ADI) \times P_i(x|\text{POD} = k) \tag{25}$$

$ADI$: an arrival/departure incident in the yard/terminal.



$P_{\text{A/DDe}}(x_{\text{tank}} \mid ADI)$: the conditional probability of derailing $x_{\text{tank}}$ tank cars given an A/D incident.

$\delta_t(j)$: 0-1 indicator; equals 1 if the car at the $j^{\text{th}}$ position in the train is a tank car, and 0 otherwise.

$L$: train length, i.e., the number of railcars in the train.

$POD(k \mid ADI)$: the probability that POD is at the $k^{\text{th}}$ position of a train given an A/D incident.

$P_i(x \mid \text{POD} = k)$: the conditional probability of derailing $x$ cars given that the POD is at $k^{\text{th}}$ position in a train on segment $i$.

Due to lower yard/terminal operating speeds relative to mainline speeds, the conditional probability of a tank car releasing given an A/D incident is reduced by multiplying a factor of 0.35 to reflect the fact that most of the yard/terminal accidents have lower severity and chances of release than mainline accidents in general, for which the base CPR factors are developed (Treichel et al., 2019). Given that $y$ tank cars derail in an A/D incident, the number of tank cars releasing contents follows a binomial distribution with $y$ independent experiments and a success probability of $0.35 \times CPR$ for each experiment. Let $P_{\text{A/DRe}}(x_{\text{tank}} \mid ADI)$ denote the probability that there are $x_{\text{tank}}$ hazmat cars releasing contents given an A/D incident in yards/terminals and $TT$ denote the total number of tank cars in a train. Once the yard- or terminal-specific derailment rates (Section 3.1.2), the number of railcars derailed (Section 3.2.1), and the number of tank cars derailed (Equation (25)) are determined, the conditional probability of releasing $x_{\text{tank}}$ tank cars can be determined as follow:



$$P_{\mathrm{A/DRe}}(x_{\mathrm{tank}}|ADI)$$

$$= \sum_{y=x_{\mathrm{tank}}}^{TT} \binom{y}{x_{\mathrm{tank}}} (0.35 \times CPR)^{x_{\mathrm{tank}}} \qquad (26)$$

$$\times (1 - 0.35 \times CPR)^{y-x_{\mathrm{tank}}} \times P_{\mathrm{A/DDe}}(y|ADI)$$

where

$ADI$: an arrival/departure incident in the yard/terminal.

$P_{\mathrm{A/DRe}}(x_{\mathrm{tank}}|ADI)$: the conditional probability that there are $x_{\mathrm{tank}}$ hazmat cars releasing contents.

given an A/D incident in yards/terminals.

$TT$: the total number of tank cars in a train.

$CPR$: the base conditional probability of release developed in Treichel et al. (2019).

$P_{\mathrm{A/DDe}}(y|ADI)$: the conditional probability that there are $y$ hazmat cars derailed in an A/D

incident.

### 3.3.3   Yard Switching Events

Liu et al. (2022) declared that calculating the number of tank cars derailed in a yard switching

incident distinguishes the yard switching approaches. Accordingly, this paper assumes that tank

cars are grouped together as a block in yard switching events. Section 3.2.2 has explained that it

is rare for manifest trains to derail more than 20 cars in a yard switching incident. Thus, in this

study, the block of tank cars (no more than 20) can be analyzed 1) as being "*switched alone*" as an

independent group of $TT$ tank cars, or 2) as being "*switched en masse*" as a block of $TT$ tank cars

behind 19 other non-hazmat cars for a total switching "cut" of $19 + TT$ railcars. The analysis

considers 19 non-hazmat railcars in front of the $TT$ tank cars because, as mentioned, the



probability of derailing more than 20 railcars in a yard switching derailment is effectively zero. Assume 20 non-tank cars followed by 20 tank cars are switched together. Within this framework, in a yard switching derailment, if the first railcar (non-tank car) of the group derails and the resulting derailment spreads back through the railcars to derail the maximum amount of 20 railcars, none of the 20 tank cars will be derailed since the final car to derail is the last non-tank car immediately in front of the first tank car in the group. In other words, when the first car to derail is more than 19 cars away from the block of $TT$ tank cars, there will be (almost) zero chance of derailing any tank cars and the scenario can be ignored (Figure 3). This paper considers the worst case (when there are at least 19 non-tank cars in front of the block of tank cars) from the conservative perspective for safety concerns. The total cut size of $TT$ tank cars using the "switched alone" approach, or $19 + TT$ railcars using the "switched en masse" approach is considered to calculate the number of hazmat cars derailed in yard switching incidents.

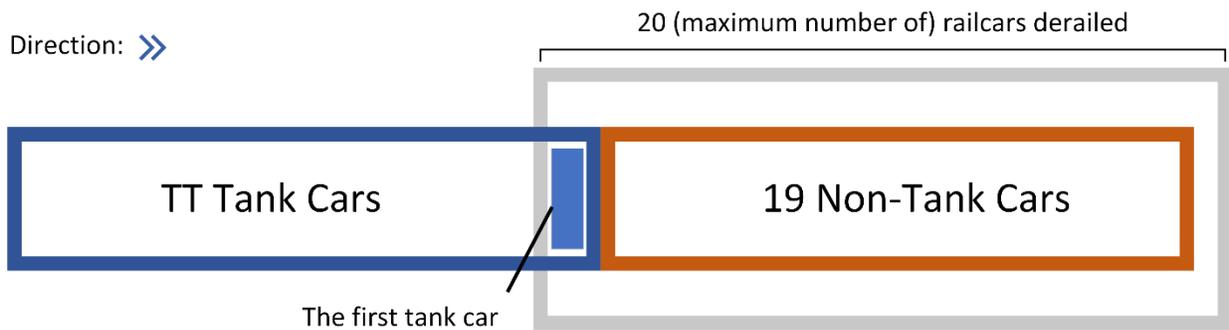

**Figure 3 Graphic explanation for the value of "19" in the "switched en masse" approach.**

Again, this paper assumes that the derailment occurs as a cut of the group of railcars switched together. Thus, for both the "switched alone" and "switched en masse" approaches, the probability



that the first car of the derailment is at $k^{\text{th}}$ position in the group of railcars switched together, when a yard switching incident has occurred, can be calculated as:

$$FCD(k|YSI) = \frac{1}{TCC} \tag{27}$$

where

$Y$: a yard switching incident.

$FCD(k|YSI)$: the probability that the first car of the derailment is at the $k^{\text{th}}$ position in the group of cars switched together given a yard switching incident.

$TCC$: the total number of cars considered in a yard switching event. For the "switched alone" approach, $TCC$ is the number of tank cars, while it is the number of tank cars plus 19 non-tank cars for the "switched en masse" approach.

For the "switched alone" approach, since all cars switched together are tank cars, the probability of derailing $x_{\text{tank}}$ tank cars given a yard switching incident can be estimated by:

$$
\begin{aligned}
P_{\text{YardDeTank}}&(x_{\text{tank}}|YSI) \\
&= \sum_{k=1}^{TCC-x_{\text{tank}}+1} FCD(k|YSI) \times P_{\text{YardDeRail}}(x_{\text{tank}}|FCD = k)
\end{aligned}
\tag{28}
$$

where

$YSI$: a yard switching incident.

$P_{\text{YardDeTank}}(x_{\text{tank}}|YSI)$: the conditional probability of derailing $x_{\text{tank}}$ tank cars given a yard switching incident.

$FCD$: the position of the first car of the derailment in the group of railcars.

$FCD(k|YSI)$: the probability that the first car of the derailment is at the $k^{\text{th}}$ position in the block of tank cars given a yard switching incident.



$TCC$: total cars considered. For the "switched alone" approach, $TCC$ is the number of tank cars.

$P_{\text{YardDeRail}}(x_{\text{tank}}|FCD = k)$: the conditional probability of derailing $x_{\text{tank}}$ railcars given that the first car of the derailment is at the $k^{\text{th}}$ position in the group of cars.

Note that in Equation (28), $k$ sums from 1 to $TCC - x_{\text{tank}} + 1$ since the remaining cases are not able to derail $x_{\text{tank}}$ tank cars.

For the "switched en masse" approach, the first car derailed can be any of 19 non-tank cars or the following block of tank cars. Thus, the probability of derailing $x_{\text{tank}}$ tank cars given a yard switching incident using the "switched en masse" approach is:

$$
\begin{aligned}
P_{\text{YardDeTank}}&(x_{\text{tank}}|YSI) \\
&= \sum_{k=x_{\text{tank}}}^{19} FCD(k|YSI) \\
&\quad \times P_{\text{YardDeRail}}(20 - k + x_{\text{tank}}|FCD = k) \\
&\quad + \sum_{k=20}^{TCC - x_{\text{tank}} + 1} FCD(k|YSI) \times P_{\text{YardDeRail}}(x_{\text{tank}}|FCD = k)
\end{aligned}
\tag{29}
$$

where

$YSI$: a yard switching incident.

$P_{\text{YardDeTank}}(x_{\text{tank}}|YSI)$: the conditional probability of derailing $x_{\text{tank}}$ tank cars given a yard switching incident.

$FCD(k|YSI)$: the probability that the first car of the derailment is at the $k^{\text{th}}$ position in the block of tank cars given a yard switching incident.



$TCC$: the total number of cars considered in a yard switching event. It is the number of tank cars plus 19 non-tank cars for the "switched en masse" approach.

$P_{\text{YardDeRail}}(x_{\text{tank}}|FCD = k)$: the conditional probability of derailing $x_{\text{tank}}$ railcars given that the first car of the derailment is at the $k^{\text{th}}$ position in the group of cars.

In Equation (29) the first term on the right-hand side calculates the probability of releasing $x_{\text{tank}}$ tank cars if the first car of the derailment is a non-tank car. The expression "$20 - k + x_{\text{tank}}$", in the term $P_{\text{YardDeRail}}(20 - k + x_{\text{tank}}|FCD = k)$, comes from the situation when the first car of the derailment is at $k^{\text{th}}$ position; the non-tank cars (from $k^{\text{th}}$ to $19^{\text{th}}$ positions) and the first $x_{\text{tank}}$ tank cars (from $20^{\text{th}}$ to $19+x_{\text{tank}}$ positions) are derailing to satisfy that there are exactly $x_{\text{tank}}$ tank cars derailing, which is the condition that there are "$19 - k + 1 + x_{\text{tank}}$" railcars derailed. The second term on the right-hand side of Equation (29) considers all cases if the first car of the derailment is a tank car. Knowing the probability distribution of the number of tank cars derailed, the probability of releasing $x_{\text{tank}}$ tank cars given a yard switching incident (denoted as $P_{\text{YardReTank}}(x_{\text{tank}}|YSI)$) can be estimated by the same method as Equation (26).

### 3.4 Number of Hazmat Cars Releasing Contents Per Shipment

Section 3.2 and Section 3.3 calculate conditional probabilities given a certain type of train derailment. This section removes the "conditions" in the probability distributions developed in Section 3.3 and calculates the probability of releasing a certain number of hazmat cars per shipment.



### 3.4.1 Line-haul Incidents on Mainlines

Section 3.1.1 defined the probability of a line-haul incident per shipment on the mainline segment $i$ as $PTD_{i,\text{main}}$, and Section 3.3.1 found that the probability of releasing $x_R$ tank cars on the mainline segment $i$ per train derailment (denoted as $P_{\text{main,i,re}}(x_R|TD)$) follows a Poisson Binomial distribution. Thus, the probability of releasing $x_R$ tank cars on the mainline segment $i$ per shipment is:

$$P_{\text{main,i,re}}(x_R) = P_{\text{main,i,re}}(x_R|TD) \times PTD_{i,\text{main}} \qquad (30)$$

where

$P_{\text{main,i,re}}(x_R)$: the probability of releasing $x_R$ tank cars on the mainline segment $i$ per shipment.

$P_{\text{main,i,re}}(x_R|TD)$: the probability of releasing $x_R$ tank cars on the mainline segment $i$ per train derailment.

$PTD_{i,\text{main}}$: the probability of a line-haul incident per shipment on the mainline segment $i$.

### 3.4.2 Unit Train Incidents in Terminals and Manifest Train Incidents in Yards

Section 3.1.2 calculated the likelihood of a train derailment per shipment during A/D events or yard switching events ($PTD_{\text{AD}}$ and $PTD_{\text{SWI}}$). To distinguish train type (unit and manifest trains), $PTD_{\text{AD}}$ is written as $PTD_{\text{AD,Unit}}$ and $PTD_{\text{AD,Manifest}}$ to represent the probability of a train derailment per shipment during A/D events in terminals for unit trains and in yards for manifest trains. Also, Section 3.3.2 and Section 3.3.3 built the probability distributions of the number of tank cars releasing contents given an A/D incident or a yard switching incident. Based on those calculations, for a unit train, the probability of releasing $x_{\text{tank}}$ tank cars per shipment in terminals is:



$$P_{\text{terminal}}(x_{\text{tank}}) = P_{\text{A/DRe}}(x_{\text{tank}}|ADI) \times PTD_{\text{ADI,Unit}} \qquad (31)$$

where

$P_{\text{terminal}}(x_{\text{tank}})$: the probability of releasing $x_{\text{tank}}$ tank cars per shipment for a unit train in terminals.

$P_{\text{A/DRe}}(x_{\text{tank}}|ADI)$: the conditional probability that there are $x_{\text{tank}}$ hazmat cars releasing contents given an A/D incident in terminal.

$PTD_{\text{ADI,Unit}}$: the probability of a train derailment per shipment during A/D events in terminals using unit trains.

In contrast, for a manifest train, the probability of releasing $x_{\text{tank}}$ tank cars per shipment in yards is:

$$P_{\text{yard}}(x_{\text{tank}}) = P_{\text{A/DRe}}(x_{\text{tank}}|ADI) \times PTD_{\text{ADI,Manifest}} \qquad (32)$$
$$+ P_{\text{YardReTank}}(x_{\text{tank}}|YSI) \times PTD_{\text{SWI}}$$

where

$P_{\text{yard}}(x_{\text{tank}})$: the probability of releasing $x_{\text{tank}}$ tank cars per shipment for a manifest train in yards.

$P_{\text{A/DRe}}(x_{\text{tank}}|ADI)$: the conditional probability that there are $x_{\text{tank}}$ hazmat cars releasing contents given an A/D incident in yards.

$PTD_{\text{ADI,Manifest}}$: the probability of a train derailment per shipment during A/D events in the yard using manifest trains.

$P_{\text{YardReTank}}(x_{\text{tank}}|YSI)$: the conditional probability that there are $x_{\text{tank}}$ hazmat cars releasing contents given a yard switching incident.

$PTD_{\text{SWI}}$: the probability of a train derailment per shipment during yard switching events.



## 3.5 Release Quantity

Using historical data from the Railway Supply Institute (RSI) and the Association of American Railroads (AAR) Tank Car Accident Database (TCAD), the RSI-AAR Railroad Tank Car Safety Research and Test Project (AAR-RSI, 2014) developed the probability distribution of release quantity from a single tank car. In this paper, the amount released from a single tank car is represented in terms of the percentage of car capacity loss based on a prior study (Treichel et al., 2019). Note that most of the non-pressure tank cars such as DOT 111s and DOT 117s have a gallon capacity of around 30,000-gallons. Table 4 presents the lading loss per car and the corresponding probability for a non-pressurized, 30,000-gallon tank car. This distribution is used to generate the amount released for all three types of risks on mainlines or in yards/terminals given the probability distributions of number of tank cars releasing contents derived from Section 3.4.

**Table 4 Probability distribution of release quantity for a single non-pressurized tank car with a gallon capacity of around 30,000-gallons. (Treichel et al., 2019)**

| Quantity of Release (QR) | Average Quantity of Release | Lading Loss per Car (gallons) | Probability |
|---|---|---|---|
| 0%-5% | 2.50% | 750 | 0.336 |
| 5%-20% | 12.50% | 3,750 | 0.095 |
| 20%-50% | 35.00% | 10,500 | 0.133 |
| 50%-80% | 65.00% | 19,500 | 0.123 |
| 80%-100% | 90.00% | 27,000 | 0.313 |



Due to information constraints, the assumption is made that the release quantity of a tank car is independent of other tank cars. Hence, for multiple tank cars releasing contents, the total release quantity is an aggregation of the release quantity from multiple tank car releases. To be more specific, the potential release quantity for a release incident (with a specific number of releasing tank cars) is the combination of the five levels in Table 4. Each incident with a particular number of tank cars releasing contents has a probability distribution of release quantity. Take, for example, a situation where it is known that 20 tank cars are releasing. In such a case, there are $5^{20}$ possible combinations of amount released, which leads to a probability distribution of the total amount of hazmat release given 20 releasing tank cars. Summing up the probability distributions of the amount released for all possible values for "the number of tank cars releasing contents," we can obtain the probability distribution of the total amount released (let $P_{re}(x)$ denote the probability of releasing $x$ gallons of contents in total from all releasing tank cars). For example, assume that there are 20 tank cars on a manifest train, and the probability of releasing 1, 2, 3, …, 20 tank cars are all identical, and equal to 0.05. Thus, there are two possible cases resulting in releasing 4,500 gallons: 1) there are six tank cars releasing contents and each of them releases 750 gallons; or 2) there are two tank cars releasing contents: one of them releases 750 gallons, and the other tank car releases 3,750 gallons (note: there is a factor "2" reflecting that there are two ways to designate which car is releasing 750 or 3,750 gallons). Thus, according to Table 4, the probability of releasing 4,500 gallons can be calculated by:



$$P(releasing\ 4{,}500\ gallons\ hazmat)$$

$$= P(there\ are\ six\ tank\ cars\ releasing\ contents)$$

$$\times P(a\ tank\ car\ releasing\ 750\ gallons)^6$$

$$+ P(there\ are\ two\ tank\ cars\ releasing\ contents) \qquad (33)$$

$$\times P(one\ tank\ car\ releasing\ 750\ gallons)$$

$$\times P(one\ tank\ car\ releasing\ 3{,}750\ gallons) \times 2$$

$$= 0.05 \times 0.336^6 + 0.05 \times 0.336 \times 0.095 \times 2 = 0.0032$$

The probability distribution of the total amount released for this 20-tank-car example is shown in Figure 4.

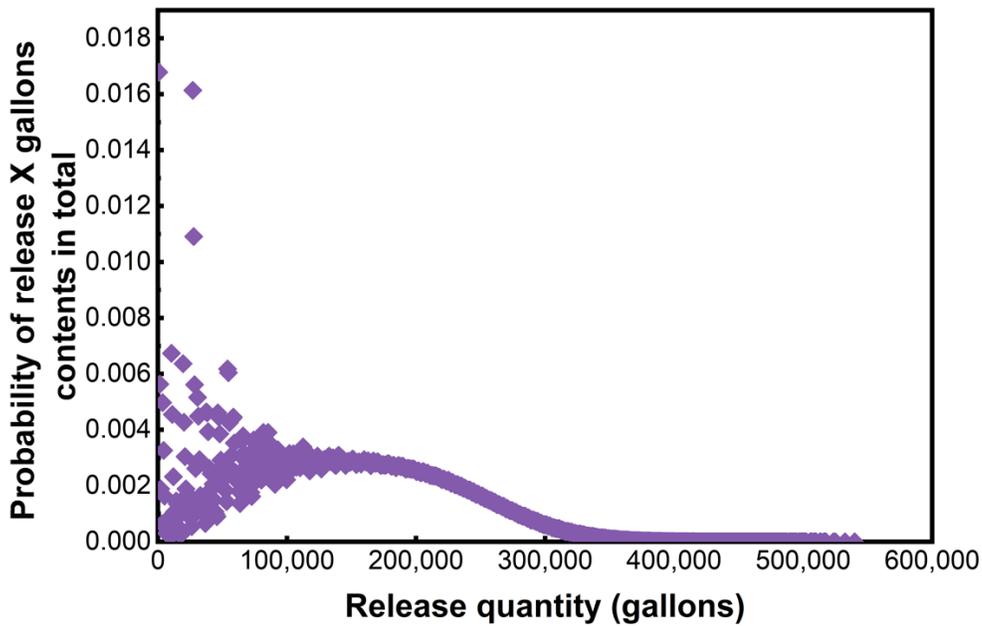

**Figure 4 Probability distribution of release quantity in gallons for the 20-tank-car example.**



### 3.6 Releasing Consequences

Performing complete consequence analyses of train operations (to injuries/fatalities for all hazmat commodities, routes, etc.) is a very significant effort. Thus, this paper reduces the scope of the problem to limit the commodities carried by tank cars to flammable liquids (crude oil and ethanol), since crude oil and ethanol combined make up a significant majority of hazmat unit train shipments. The consequence analyses performed in this section demonstrate the approach to analyzing consequences resulting from shipments in unit trains versus manifest trains. One difficulty in performing consequence analyses is that the results are often controlled by the most severe events which are extremely rare, and the methodology makes it difficult to include consequences from conditions that have not been previously observed. A consequence analysis for rail transportation of flammable liquids performed prior to 2013 would likely not have considered that an event like the Lac-Mégantic rail disaster, which resulted in 47 fatalities and more than 30 buildings destroyed, was possible. One such type of severe consequence that has not been significantly considered for flammable liquids by rail is an uncontrolled fire spread.

This paper leverages the Hazard Prediction and Assessment Capability (HPAC) with its associated analysis modules and the Nuclear Capabilities Services (NuCS) framework to assess consequences of industrial accidents (e.g., hazmat spill fires). The proposed approach applies the HPAC and NuCS toolsets to analyze a series of derailment events (flammable liquid releases) at representative real-world locations with varying population densities and various release sizes. Three representative locations along a rail line (urban, suburban, and rural) in the NuCS database are selected for the derailment sites.



The HPAC tool contains an OILSPILL model (a spreadsheet-based tool) that predicts the area and volume of contained and uncontained crude oil spills at selected locations with varying population densities and using various release sizes. The tool is geo-referenced and imports building, vegetation, and population data based on an input location. The amount of oil spilled can be estimated using railcar volumes, and this paper assumes that all fuel in the spill footprint will ignite for the conservative consideration. Then, the fuel spill distribution is mapped on the ground into the fuel files for the fire spread/casualty code (QUIC-FST) (Crepeau & Etheridge, 2019; Etheridge, 2020). The fire spread code is then run to provide a time-dependent map of the fuel consumed by fire, and the fire casualty model provides a time-dependent map of the casualties (fatalities and injuries) due to the propagating fire with a breakdown of casualties.

The fire spread/casualty code can be applied to a given region with a defined population to provide an estimate of fire casualties for that location and population. It calculates the probability of injury and fatality due to a thermal dose in each computational cell in the scene. Applying a random number generator and the probabilities to the population density, it arrives at an estimate of casualties (combined injuries and fatalities) for each computational cell. For a given spill, casualties are dependent on the vegetation and building distribution in the area.

Section 3.5 has built the probability distribution of release quantity for a unit train carrying 100 tank cars and a manifest train with a block of 20 tank cars. The results found that the probability of releasing more than 150,000 gallons of content in total is almost zero. Thus, for the consequence model in this paper, we focus on the total casualties caused by one, three, or five tank cars releasing contents, which represent small (30,000 gallons), medium (90,000 gallons), and large sizes



(150,000 gallons) of tank car release incidents, respectively. By performing a series of analyses with the above tool at a series of selected locations, we can develop a set of consequence curves (Figure 5) for casualties as a function of evacuation time with characterization values set in Table 5.

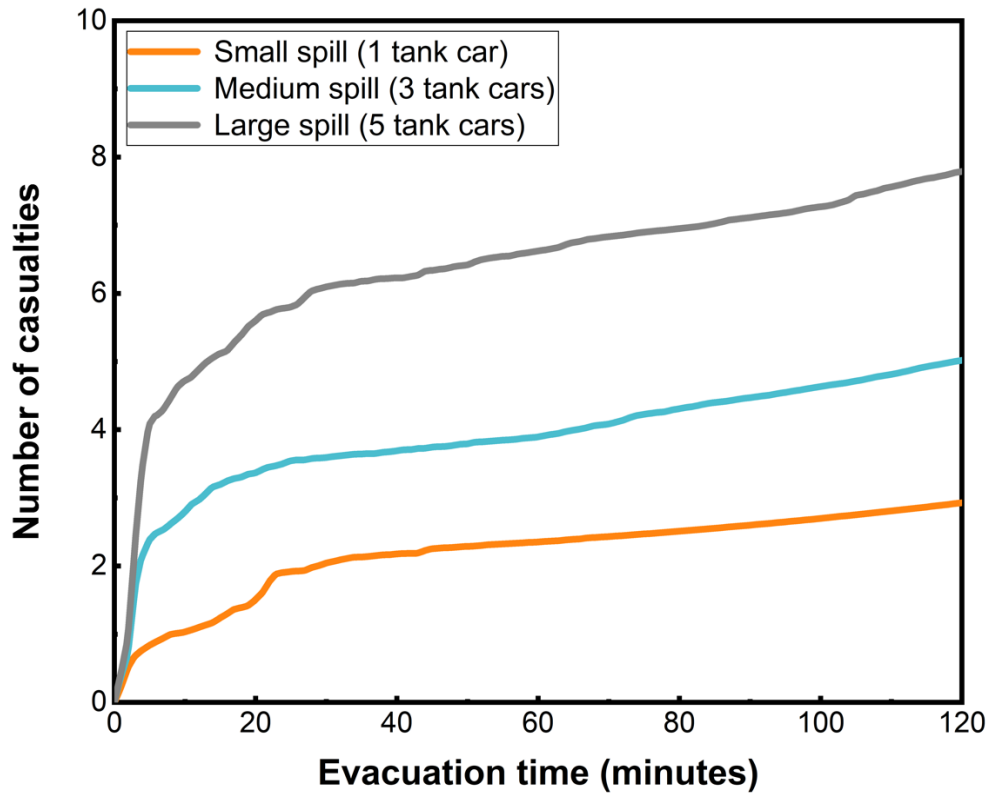

**Figure 5 Casualties from fire spread in the QUIC-FST analyses**

**Table 5 Values set on calculating casualties**

| Route characterization | |
| --- | --- |
| Urban track percentage | 1% |
| Suburban track percentage | 4% |
| Rural track percentage | 95% |
| **Weather characterization** | |
| Low wind percentage | 50% |
| Medium wind percentage | 49% |
| High wind percentage | 1% |
| **Evacuation time** | |
| Nearby building evacuation time | 4 mins |
| Evacuation time for hazard zone | 120 mins |

We assume that there are no casualties when no tank cars release. Thus, we can piecewise-linearly interpolate casualties when the release quantity is between 0-30,000 gallons, 30,000-90,000 gallons, and 90,000-150,000 gallons. Equation (34) is the formula to calculate expected total casualties at a specific evacuation time (it applies to all three types of incidents).

$$TC(t) = \sum_{0 < x \leq 150,000} P_{\text{re}}(x) \times C(x,t) \tag{34}$$

where:

$TC(t)$: the total casualties after $t$ minutes caused by a releasing incident.

$P_{\text{re}}(x)$: the probability of releasing $x$ gallons of contents in total from all releasing tank cars (from Section 3.5).

$C(x,t)$: the expected total casualties caused by releasing $x$ gallons of content at response time $t$, and $t \in [0,120]$ in minutes (Figure 5).

Note that Equation (34) applies to all three types of derailments: the line-haul (mainline) train derailment, the A/D train derailment, and the yard switching train derailment. To distinguish train



types and derailment types, $TC(t)$ is written as $TC_{\text{main,i,Unit}}(t)$ or $TC_{\text{main,i,Manifest}}(t)$ to represent total casualties per shipment on the mainline segment $i$ using a unit train or a manifest train, $TC_{\text{ADI,Unit}}(t)$ to represent total casualties per shipment during A/D events using a unit train, and $TC_{\text{Yard}}(t)$ to represent total casualties per shipment considering both A/D events and yard switching events using a manifest train.

### *3.7 Summary*

According to the above descriptions, the total expected casualties per train derailment caused by a flammable liquid release can be calculated following the event chain described in Sections 3.1 to 3.6. Let the operator $\lceil x \rceil$ be the smallest integer greater than or equal to $x$. If there are $\delta$ tank cars that need to be transported, the number of shipments using unit trains (each unit train can carry $c_{\text{unit}}$ tank cars) is $\left\lceil \frac{\delta}{c_{\text{unit}}} \right\rceil$, and the total expected casualties per traffic demand can be calculated by:

$$TC_{\text{Final}}(t) = \left( \sum_{\forall i} TC_{\text{main,i,Unit}}(t) \times L_i + TC_{\text{ADI,Unit}}(t) \right) \times \left\lceil \frac{\delta}{c_{\text{unit}}} \right\rceil \qquad (35)$$

where:

$TC_{\text{Final}}(t)$: total casualties if the evacuation process starts after $t$ minutes of a releasing incident per traffic demand.

$\delta$: the number of tank cars that need to be transported.

$L_i$: the length (in miles) of track segment $i$.

$c_{\text{unit}}$: the capacity of the unit train, i.e., the number of tank cars a unit train can carry.

$TC_{\text{main,i,Unit}}(t)$: total casualties per shipment on the mainline segment $i$ using a unit train (obtained by applying Equation (34) in Section 3.6).



$TC_{\text{ADI,Unit}}(t)$: total casualties per shipment during A/D events using a unit train (obtained by applying Equation (34) in Section 3.6).

Using manifest trains (each manifest train can carry $c_{\text{manifest}}$ tank cars) to perform the same service, the number of shipments needed is $\left\lceil \frac{\delta}{c_{\text{manifest}}} \right\rceil$, and the total casualties per traffic demand can be estimated by:

$$TC_{\text{Final}}(t) = \left( \sum_{\forall i} TC_{\text{main,i,Manifest}}(t) \times L_i + TC_{\text{Yard}}(t) \right) \times \left\lceil \frac{\delta}{c_{\text{manifest}}} \right\rceil \qquad (36)$$

where:

$TC_{\text{Final}}(t)$: total casualties if the evacuation process starts after $t$ minutes of a releasing incident per traffic demand.

$\delta$: the number of tank cars that need to be transported.

$L_i$: the length (in miles) of track segment $i$.

$c_{\text{manifest}}$: the capacity of the manifest train, i.e., the number of tank cars a manifest train can carry.

$TC_{\text{main,i,Manifest}}(t)$: total casualties per shipment on the mainline segment $i$ using a manifest train (obtained by applying Equation (34) in Section 3.6).

$TC_{\text{Yard}}(t)$: total casualties per shipment during both A/D events and yard switching events using a manifest train (obtained by applying Equation (34) in Section 3.6).

## 4. Concluding Remarks

A large amount of effort has been made to mitigate the risk related to rail transportation of hazardous materials due to its devastating consequences. However, relatively limited prior



research compared different service options (unit trains versus manifest trains) considering both mainline and yard risk components. This paper proposes a novel methodology quantifying the total risks as expected casualties for any service options knowing train configuration, tank car placement, yard type, and switching approach. There are two (three) types of risks that a unit train (manifest train) encounters per shipment. A unit train experiences arrival/departure risks in terminals and line-haul risks on mainlines, while a manifest train has additional risks during yard switching events. For each of these risks, multiple probabilistic models are built to conduct a comprehensive risk analysis of transporting hazmat by unit trains versus manifest trains. A variety of parameters are estimated for the unit train and the manifest train separately using historical derailment data for 1996-2018, considering the differences associated with various service options.

This comprehensive study is separated into Part I (methodology) and Part II (a practical case study). In the part II paper, a case study is conducted considering different levels of tank car positions, classification yard types, yard switching approaches, and train configurations. It compares unit trains and manifest trains by applying the methodology proposed in this paper. More practical insights will be discussed in the part II paper.

## Acknowledgement

This research was funded through a contract by the Federal Railroad Administration (693JJ619C000017). However, all views, analyses, and errors are solely of the authors.



## Appendix A

**Table A.1 FRA-reportable Class I mainline train derailment data, 1996-2018**

### (a) Unit train derailments

| Cause group | Frequency | Percent of total |
|---|---|---|
| Broken Rails or Welds | 440 | 17.87 |
| Broken Wheels (Car) | 230 | 9.34 |
| Bearing Failure (Car) | 182 | 7.39 |
| Buckled Track | 152 | 6.17 |
| Other Axle/Journal Defects (Car) | 152 | 6.17 |
| Track Geometry (excl. Wide Gauge) | 141 | 5.73 |
| Obstructions | 98 | 3.98 |
| Wide Gauge | 87 | 3.53 |
| Roadbed Defects | 71 | 2.88 |
| Other Wheel Defects (Car) | 70 | 2.84 |
| Turnout Defects - Switches | 65 | 2.64 |
| Track-Train Interaction | 58 | 2.36 |
| Other Miscellaneous | 56 | 2.27 |
| Misc. Track and Structure Defects | 50 | 2.03 |
| Lading Problems | 46 | 1.87 |
| Joint Bar Defects | 46 | 1.87 |
| Coupler Defects (Car) | 41 | 1.67 |
| Other Rail and Joint Defects | 40 | 1.62 |
| Use of Switches | 38 | 1.54 |
| Sidebearing, Suspension Defects (Car) | 36 | 1.46 |
| Train Handling (excl. Brakes) | 32 | 1.3 |
| Non-Traffic, Weather Causes | 31 | 1.26 |
| Rail Defects at Bolted Joint | 30 | 1.22 |
| Train Speed | 28 | 1.14 |
| Truck Structure Defects (Car) | 27 | 1.1 |
| Centerplate/Carbody Defects (Car) | 22 | 0.89 |
| All Other Car Defects | 22 | 0.89 |
| Misc. Human Factors | 21 | 0.85 |
| Stiff Truck (Car) | 15 | 0.61 |
| Switching Rules | 15 | 0.61 |
| Failure to Obey/Display Signals | 14 | 0.57 |
| Other Brake Defect (Car) | 14 | 0.57 |
| Handbrake Operations | 12 | 0.49 |
| Brake Rigging Defect (Car) | 12 | 0.49 |
| Loco Electrical and Fires | 11 | 0.45 |
| Track/Train Interaction (Hunting) (Car) | 10 | 0.41 |
| Brake Operation (Main Line) | 9 | 0.37 |



| Cause group | Frequency | Percent of total |
|---|---|---|
| Mainline Rules | 9 | 0.37 |
| Signal Failures | 8 | 0.32 |
| Loco Trucks/Bearings/Wheels | 8 | 0.32 |
| Turnout Defects - Frogs | 5 | 0.2 |
| All Other Locomotive Defects | 3 | 0.12 |
| Brake Operations (Other) | 2 | 0.08 |
| UDE (Car or Loco) | 1 | 0.04 |
| Employee Physical Condition | 1 | 0.04 |
| Air Hose Defect (Car) | 1 | 0.04 |
| Total | 2,462 | 100 |

**(b) Manifest train derailments**

| Cause group | Frequency | Percent of total |
|---|---|---|
| Broken Rails or Welds | 639 | 11.59 |
| Track Geometry (excl. Wide Gauge) | 391 | 7.09 |
| Bearing Failure (Car) | 343 | 6.22 |
| Train Handling (excl. Brakes) | 324 | 5.88 |
| Obstructions | 243 | 4.41 |
| Track-Train Interaction | 212 | 3.84 |
| Lading Problems | 211 | 3.83 |
| Wide Gauge | 186 | 3.37 |
| Coupler Defects (Car) | 184 | 3.34 |
| Use of Switches | 182 | 3.30 |
| Broken Wheels (Car) | 173 | 3.14 |
| Sidebearing, Suspension Defects (Car) | 164 | 2.97 |
| Other Wheel Defects (Car) | 164 | 2.97 |
| Brake Operation (Main Line) | 163 | 2.96 |
| Centerplate/Carbody Defects (Car) | 148 | 2.68 |
| Buckled Track | 147 | 2.67 |
| Other Miscellaneous | 145 | 2.63 |
| Turnout Defects - Switches | 142 | 2.58 |
| Misc. Track and Structure Defects | 98 | 1.78 |
| Train Speed | 94 | 1.70 |
| Stiff Truck (Car) | 85 | 1.54 |
| Roadbed Defects | 82 | 1.49 |
| Joint Bar Defects | 70 | 1.27 |
| Other Axle/Journal Defects (Car) | 64 | 1.16 |
| Other Brake Defect (Car) | 64 | 1.16 |
| Loco Trucks/Bearings/Wheels | 63 | 1.14 |
| All Other Car Defects | 62 | 1.12 |
| Track/Train Interaction (Hunting) (Car) | 58 | 1.05 |
| Misc. Human Factors | 58 | 1.05 |
| Switching Rules | 55 | 1.00 |
| Other Rail and Joint Defects | 51 | 0.92 |
| Rail Defects at Bolted Joint | 51 | 0.92 |
| Handbrake Operations | 49 | 0.89 |
| Non-Traffic, Weather Causes | 44 | 0.80 |
| Failure to Obey/Display Signals | 39 | 0.71 |
| Brake Rigging Defect (Car) | 35 | 0.63 |
| All Other Locomotive Defects | 35 | 0.63 |
| Signal Failures | 35 | 0.63 |
| Air Hose Defect (Car) | 33 | 0.60 |
| Truck Structure Defects (Car) | 25 | 0.45 |
| Loco Electrical and Fires | 23 | 0.42 |
| Mainline Rules | 23 | 0.42 |
| Turnout Defects - Frogs | 20 | 0.36 |
| Radio Communications Error | 12 | 0.22 |
| UDE (Car or Loco) | 10 | 0.18 |
| Brake Operations (Other) | 6 | 0.11 |
| TOFC/COFC Defects | 5 | 0.09 |
| Employee Physical Condition | 2 | 0.04 |
| Handbrake Defects (Car) | 2 | 0.04 |
| Total | 5,514 | 100 |

## Appendix B

### Table B.1 FRA-reportable Class I yard train arrival/departure event derailment data, 1996-2018  (Zhao & Dick, 2022)

#### (a) Unit train derailments

| Cause group | Frequency | Percent of total |
|---|---|---|
| Broken rails or welds | 224 | 26.79 |
| Wide gauge | 106 | 12.68 |
| Turnout defects: switches | 105 | 12.56 |
| Use of switches | 79 | 9.45 |
| Switching rules | 42 | 5.02 |
| Miscellaneous track and structure defects | 29 | 3.47 |
| Track geometry (excluding wide gauge) | 27 | 3.23 |
| Other miscellaneous | 26 | 3.11 |
| Other wheel defects (car) | 19 | 2.27 |
| Roadbed defects | 18 | 2.15 |
| Rail defects at bolted joint | 13 | 1.56 |
| Train handling (excluding brakes) | 13 | 1.56 |
| Train speed | 12 | 1.44 |
| Stiff truck (car) | 12 | 1.44 |
| Lading problems | 11 | 1.32 |
| Other rail and joint defects | 10 | 1.20 |
| Track–train interaction | 9 | 1.08 |
| Side bearing and suspension defects (car) | 8 | 0.96 |
| Miscellaneous human factors | 8 | 0.96 |
| Handbrake operations | 7 | 0.84 |
| Joint bar defects | 7 | 0.84 |
| Buckled track | 6 | 0.72 |
| Signal failures | 5 | 0.60 |
| Nontraffic, weather causes | 5 | 0.60 |
| Brake rigging defect (car) | 4 | 0.48 |
| Failure to obey or display signals | 3 | 0.36 |
| Locomotive trucks, bearings, and wheels | 3 | 0.36 |
| All other locomotive defects | 3 | 0.36 |
| All other car defects | 3 | 0.36 |
| Brake operation (main line) | 2 | 0.24 |
| Centerplate or car body defects (car) | 2 | 0.24 |
| Extreme weather | 2 | 0.24 |
| Bearing failure (car) | 2 | 0.24 |
| Turnout defects: frogs | 2 | 0.24 |
| Broken wheels (car) | 2 | 0.24 |
| Locomotive electrical and fires | 2 | 0.24 |
| Handbrake defects (car) | 1 | 0.12 |
| Brake operations (other) | 1 | 0.12 |
| UDE (car or locomotive) | 1 | 0.12 |
| Other brake defect (car) | 1 | 0.12 |
| Mainline rules | 1 | 0.12 |
| Total | 836 | 100 |

## (b) Manifest train derailments

| Cause group | Frequency | Percent of total |
|---|---|---|
| Switching rules | 908 | 15.45 |
| Use of switches | 766 | 13.03 |
| Broken rails or welds | 685 | 11.66 |
| Wide gauge | 625 | 10.63 |
| Turnout defects: switches | 486 | 8.27 |
| Train handling (excluding brakes) | 407 | 6.93 |
| Other miscellaneous | 206 | 3.51 |
| Handbrake operations | 195 | 3.32 |
| Train speed | 183 | 3.11 |
| Miscellaneous track and structure defects | 155 | 2.64 |
| Track–train interaction | 150 | 2.55 |
| Track geometry (excluding wide gauge) | 141 | 2.40 |
| Brake operation (main line) | 136 | 2.31 |
| Lading problems | 79 | 1.34 |
| Other wheel defects (car) | 70 | 1.19 |
| Signal failures | 63 | 1.07 |
| Side bearing and suspension defects (car) | 59 | 1.00 |
| Coupler defects (car) | 56 | 0.95 |
| Stiff truck (car) | 54 | 0.92 |
| Roadbed defects | 51 | 0.87 |
| Radio communications error | 48 | 0.82 |
| Rail defects at bolted joint | 37 | 0.63 |
| Miscellaneous human factors | 37 | 0.63 |
| Centerplate or car body defects (car) | 28 | 0.48 |
| Turnout defects: frogs | 28 | 0.48 |
| Mainline rules | 26 | 0.44 |
| Nontraffic, weather causes | 22 | 0.37 |
| All other car defects | 18 | 0.31 |
| Other rail and joint defects | 16 | 0.27 |
| Brake operations (other) | 15 | 0.26 |
| Other brake defect (car) | 15 | 0.26 |
| Brake rigging defect (car) | 14 | 0.24 |
| Extreme weather | 13 | 0.22 |
| Locomotive trucks, bearings, and wheels | 12 | 0.20 |
| All other locomotive defects | 12 | 0.20 |
| Buckled track | 11 | 0.19 |
| Failure to obey or display signals | 10 | 0.17 |
| Joint bar defects | 10 | 0.17 |
| Broken wheels (car) | 8 | 0.14 |
| Obstructions | 4 | 0.07 |
| Handbrake defects (car) | 4 | 0.07 |
| Employee physical condition | 4 | 0.07 |
| Truck structure defects (car) | 4 | 0.07 |
| Air hose defect (car) | 2 | 0.03 |
| UDE (car or locomotive) | 1 | 0.02 |
| Bearing failure (car) | 1 | 0.02 |
| Locomotive electrical and fires | 1 | 0.02 |
| Track–train interaction (hunting) (car) | 1 | 0.02 |
| Total | 5,877 | 100 |